\documentclass[]{IEEEtran}

\usepackage{amsmath}
\usepackage{amssymb}
\usepackage{amsthm}
\usepackage{tikz}
\usepackage{pgfplots}
\newtheorem{proposition}{Proposition}
\theoremstyle{definition}
\newtheorem*{remark}{Remark}

\usetikzlibrary{dsp, chains}
\usetikzlibrary{plotmarks}
\pgfplotsset{compat=newest} 
\pgfplotsset{plot coordinates/math parser=false}
\newlength\figurewidth
\title{Mixing oscillators for phase noise reduction}
\author{\IEEEauthorblockN{Paul Ferrand,~\IEEEmembership{Member,~IEEE}} \thanks{The author is with the Mathematical and Algorithmic Sciences Laboratory, Huawei Technologies France, 92100 Boulogne-Billancourt, France. Email: paul.ferrand@huawei.com.}}

\definecolor{col1}{rgb}{0.0000,0.4470,0.7410}%
\definecolor{col2}{rgb}{0.8500,0.3250,0.0980}%
\definecolor{col3}{rgb}{0.9290,0.6940,0.1250}%
\definecolor{col4}{rgb}{0.4940,0.1840,0.5560}%
\definecolor{col5}{rgb}{0.4660,0.6740,0.1880}%
\definecolor{col6}{rgb}{0.3010,0.7450,0.9330}%

\begin{document}
\maketitle

\begin{abstract}
The output of oscillators is usually not stable over time.
In particular, phase variations---or \emph{phase noise}---corrupts the oscillations.
In this letter, we describe a circuit that designed to average the phase noise processes and frequency offsets in frequency-matched oscillators.
The basic circuit uses the independence of 2 phase noise processes to provide a cleaner oscillating output with lower phase noise variance.
We describe extensions of the circuit designed to average out more than 2 oscillators, as well as a single one through delay elements.
In all the examples covered, we provide a theoretical analysis of the resulting phase noise process when the input phase noise processes follow a Wiener model.
\end{abstract}
\section{Introduction}
\label{sec:intro}
Noise effects can perturb not only the amplitude but also the phase of local oscillators.
Industrial progress lead to an increased signal-to-noise ratio (SNR) in most circuit components \cite{lee2000b, voicu2013}.
As improvements in SNR reduce only the amplitude noise, phase noise processes dominate the analog impairements of local oscillators and become one of the major non-idealities to consider in communication systems.
They corrupt the separation of frequency bins in orthogonal frequency division multiplexing (OFDM) signals and induce complex inter-carrier interference \cite{tomba1998}.
They impact the estimation of channel realizations \cite{mehrpouyan2012} and beamforming techniques \cite{hhne2010} in multiple input, multiple output (MIMO) applications.
The relative prominence of phase noise effects also increases with the oscillating frequency \cite{voicu2013}---a major drawback for mmWave applications~\cite{zhang2015}.

In this letter, we detail a circuit that can improve the phase noise performance of any electronic oscillator.
The circuit averages independent or slightly correlated phase noise effects in order to reduce their variance over time, and thus stabilizes the oscillator output.
The oscillating signal undergoes one or more mixing steps to sum the phase noise processes, and a frequency dividing step to extract their average.
Llopis \emph{et al.} showed that getting a high frequency oscillator through a frequency dividing circuit could provide up to a 6dB reduction in phase noise power spectral density (PSD) over a free running oscillator \cite{llopis1993}.
The present analysis uses the same mathematical mechanism to clean the output of frequency-matched oscillators.
We present the basic circuit in Sec.\ref{sec:base_circuit} and analyze its theoretical effect on the phase noise statistics.
We then show in Sec.\ref{sec:ext} that the circuit may be extended to more oscillators. 
It can also be adapted to average the phase noise effects using a single oscillator and an analog delay line.
We derive the PSD of the phase noise process in the latter case, and analyze the results.

In the following, let $\mathbb E[\cdot]$ denote the expectation of a random variable. Let $\delta(\cdot)$ be the Dirac delta function. Let $\bar z$ denote the conjugate of the complex number $z$. We use the notation $x_t$ to denote a continuous-time random process with time variable $t$; further indexing of the random process is denoted using a superscript, as in $x_t^i$.
\section{System model}
\label{sec:model}
We consider oscillators indexed by $i$, all built to output a sinusoid with a common nominal frequency $f_c$. Each oscillator presents a frequency offset $f_i$  and a zero-mean phase process $\theta_t^{i}$. Let $\omega_i = 2 \pi (f_c + f_i)$. The output of the $i$\textsuperscript{th} oscillator is thus
\begin{equation}
	s^i_t = \cos\left(\omega_i t + \theta_t^{i}\right).
	\label{eq:osc}
\end{equation}
Note that we normalize the output power of the basic oscillators, and neglect the amplitude noise.
We assume the following:
\begin{itemize}
	\item The frequency offset is randomly distributed with finite variance $\sigma^2$.
	\item The phase process $\theta_t^i$ is a real Gaussian process wrapped on the circle. It is modeled as
	\begin{align}
		\theta_t^i &= \theta_0^i + \int_0^t w^i_\tau d\tau \mod 2\pi
	\end{align}
	where $w_t^i$ is a white gaussian noise process with mean $\mathbb E[w_t^i] = 0$ and autocorrelation $\mathbb E[w_{t_1}^i w_{t_2}^i] = 2 \pi \beta \delta(t_2 - t_1)$.
	Such a process is called a Wiener process; one can show that $\mathbb E[\theta_t^i] = \theta_0^i$ and $\mathbb E[(\theta_{t_1}^i-\theta_0^i)(\theta_{t_2}^i-\theta_0^i)] = 2 \pi \beta \min(t_1,t_2)$.
\end{itemize}
In communication applications, we're interested in the phase shift $u^i_t =\exp\left(\jmath \theta_t^i\right)$. The autocorrelation of the phase shift is \cite{lee2000b}
\begin{align}
	R_u(t,t + \tau) = \mathbb E\left[u^i_t \overline{u^i_{t+\tau}}\right] = \exp\left(- \pi \beta|\tau|\right).
\end{align}
The phase shift is thus a stationary process---its autocorrelation is independent of the time variable $t$. For completeness, note that the PSD of the phase shift process is the Fourier Transform of its autocorrelation; it can be derived as the so-called Lorentzian \cite{lee2000b}
\begin{equation}
	S_u(\omega) = \frac{\pi\beta}{(\pi\beta/2)^2 + \omega^2} \qquad \omega=2\pi f.
	\label{eq:psd}
\end{equation}
In this formulation, $f$ is the frequency offset from the carrier.

\section{A mixing circuit to average oscillators}
\label{sec:base_circuit}
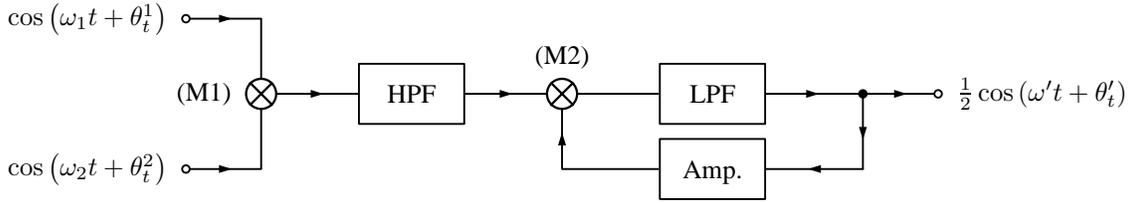
\begin{figure*}
\centering
\begin{tikzpicture}[transform shape, scale=1]
	\node[dspnodeopen,dsp/label=left] (osc1) at (0,1) {$\cos\left(\omega_1 t + \theta_t^1\right)$};
	\node[coordinate] (cosc1) at (1,1) {};
	\node[coordinate] (cosc2) at (1,-1) {};
	\node[dspnodeopen,dsp/label=left] (osc2) at (0,-1) {$\cos\left(\omega_2 t + \theta_t^2\right)$};
	\node[dspmixer,dsp/label=left] (mul) at (1,0) {(M1)};
	\node[dspfilter] (hi_filt) at (3,0) {HPF};
	\node[dspmixer,dsp/label=above] (mul2) at (5,0) {(M2)};
	\node[dspfilter] (low_filt) at (7,0) {LPF};
	\node[dspfilter] (amp) at (7,-1) {Amp.};
	\node[dspnodefull] (splitter) at (9,0) {};
	\node[dspnodeopen,dsp/label=right] (cout) at (10,0) {$\frac{1}{2}\cos\left( \omega' t + \theta_t'\right)$};
	\node[coordinate] (cfb1) at (9,-1) {};
	\node[coordinate] (cfb2) at (5,-1) {};
	\begin{scope}[start chain]
		\chainin (osc1);
		\chainin (cosc1) [join=by dspflow];
		\chainin (mul) [join=by dspline];
	\end{scope}
	\begin{scope}[start chain]
	
		\chainin (osc2);
		\chainin (cosc2) [join=by dspflow];
		\chainin (mul) [join=by dspline];
		\chainin (hi_filt) [join=by dspflow];
		\chainin (mul2) [join=by dspflow];
		\chainin (low_filt) [join=by dspline];
		\chainin (splitter) [join=by dspflow];
		\chainin (cout) [join=by dspflow];
	\end{scope}
	\begin{scope}[start chain]
		\chainin (splitter);
		\chainin (cfb1) [join=by dspflow];
		\chainin (amp) [join=by dspflow];
		\chainin (cfb2) [join=by dspline];
		\chainin (mul2) [join=by dspflow];
	\end{scope}
\end{tikzpicture}
\caption{The averaging circuit. ``HPF'' and ``LPF'' denote a high-pass and low-pass filter respectively, and ``Amp.'' an amplifier. At the steady state, we have an oscillating output with angular frequency $\omega' = (\omega_1 + \omega_2)/2$ and phase noise process $\theta_t' = (\theta_t^1 + \theta_t^2)/2$.}
\label{fig:circuit}
\end{figure*}

Independent oscillators will exhibit independent phase noise and frequency offsets.
These effects are centered on their nominal frequency $f_c$ as per \eqref{eq:osc}.
We will use their independence to reduce their variance through averaging.

\subsection{Description of the circuit}
Consider the circuit represented on Fig.\ref{fig:circuit}.
At the output of the first mixer (M1), the signal $s^1_t$ is
\begin{equation}
\begin{split}
	s^1_t = &\frac{1}{2}\cos\bigg((\omega_1 + \omega_2) t + (\theta_t^1 + \theta_t^2)\bigg) \\
		&+ \frac{1}{2}\cos\bigg((\omega_1 - \omega_2) t + (\theta_t^1 - \theta_t^2)\bigg).
\end{split}
\end{equation}
The output of the high-pass filter thus removes the low frequency signal oscillating around $(f_1 - f_2)$ and keeps the signal oscillating around $(f_1 + f_2)$.
Beginning at the second mixer (M2) is a frequency dividing circuit \cite{Miller1939} that moves the signal back to the nominal oscillator frequency $f_c$.
At the steady state, the output of the second mixer is
\begin{equation}
\begin{split}
	s^2_t = &\frac{1}{4}\cos\bigg(\omega't+\theta_t'\bigg)\cos\bigg((\omega_1 + \omega_2) t + (\theta_t^1 + \theta_t^2)\bigg)\\
	 = &\frac{1}{8}\cos\bigg((\omega_1 + \omega_2 + \omega') t + (\theta_t^1 + \theta_t^2+\theta')\bigg) \\
		&+ \frac{1}{8}\cos\bigg((\omega_1 + \omega_2 - \omega') t + (\theta_t^1 + \theta_t^2 - \theta_t')\bigg).
\end{split}
\end{equation}
The signal goes through a lowpass filter and an amplifier scaling it back to half the nominal power output of the oscillators.
Overall, the steady state equation for the system
\begin{equation}
	\omega't+\theta_t' = (\omega_1 + \omega_2 - \omega') t + (\theta_t^1 + \theta_t^2 - \theta_t'),
\end{equation}
from which we deduce
\begin{equation}
	\theta_t' = \frac{\theta_t^1 + \theta_t^2}{2} \text{ and } \omega' = \frac{\omega_1 + \omega_2}{2}. \label{eq:avc_2osc}
\end{equation}
The highpass filter ramp does not have to be very sharp---in essence, it has to discriminate between the $|f_1 + f_2| \approx 2 f_c$ and $|f_1 - f_2| \approx 0$.
The value of $f_c$ is typically large, in the order of 1--100 GHz is most communication applications.
Similarly, the lowpass filter has to discriminate between $4f_c$ and $f_c$.
Regenerative dividers are doable with a very low amount of phase noise \cite{Ferre-Pikal1999}.
Assuming good mixers and high-pass filters, the overall phase noise added by the circuit should be negligible with respect to the noise processes of the original oscillators.

\subsection{Theoretical analysis}

As seen in \eqref{eq:avc_2osc}, the frequency offset of the resulting signal is the average of the frequency offsets of the original oscillators.
Assuming these frequency offsets are drawn from a random process with mean $f_c$ variance $\sigma^2$, the resulting process has the same mean and half the variance. 
The resulting variable frequency offset thus has variance $\sigma^2/2$ in our setting. If the frequency offsets are uniformly distributed in $f_c \pm f_o$, the resulting random variable follows a Bates distribution with $n=2$ on the same support \cite{bates1955a}.
If they are assumed normally distributed, the resulting random variable is also normal with half the variance.

We make a similar argument for the phase noise process.
The mean of the resulting phase noise process can be computed as
\begin{align}
	\mathbb E[\theta_t'] &= \mathbb E\left[\theta_t^1\right]/2 + \mathbb E\left[\theta_t^2\right]/2 = \frac{\theta_0^1+\theta_0^2}{2} = \theta_0'
\end{align} 
and the process auto-correlation is
\begin{align*}
	&\mathbb E\left[(\theta_{t_1}' - \theta_0')(\theta_{t_2}'-\theta_0')\right] = \notag\\
	\frac{1}{4}\bigg(&\mathbb E\left[(\theta_{t_1}^1- \theta_0^1 )(\theta_{t_2}^1- \theta_0^1)\right] + \mathbb E\left[(\theta_{t_1}^1 - \theta_0^1) (\theta_{t_2}^2-\theta_0^2)\right] \notag\\
	+ &\mathbb E\left[(\theta_{t_1}^2- \theta_0^2) (\theta_{t_2}^1- \theta_0^1)\right] + \mathbb E\left[(\theta_{t_1}^2 - \theta_0^2) (\theta_{t_2}^2- \theta_0^2)\right]\bigg)
\end{align*} 
Since the original phase noise processes are independent, the cross-correlation terms are null and the resulting process has half the auto-correlation of the original one.
In all performance measures, this translates into a 3 dB reduction in the noise variance and improves the phase stability of the oscillators.
Note at this point that the variance reducing effect is not conditioned on any particular phase noise model and only relies on independence---or rather lack of correlation---between the phase noise processes.
For processes following the Wiener model of Sec.\ref{sec:model}, the auto-correlation of the resulting process is thus
\begin{equation}
	\mathbb E\left[(\theta_{t_1}' - \theta_0')(\theta_{t_2}'-\theta_0')\right] = \pi\beta \min(t_1,t_2).
\end{equation}

\section{Extending the basic circuit}
\label{sec:ext}
The basic oscillator averaging technique can be applied in other configurations. In this section we show how to average more than 2 oscillators by expanding the mixing stage and going through an $n$-frequency divider. We also discuss the possibility of averaging the oscillator with a delayed version of itself. 

\subsection{Averaging an arbitrary number of oscillators}
\label{sec:n_osc}
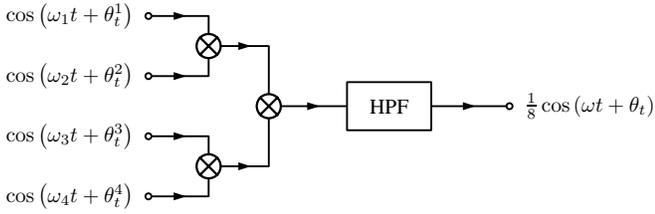
\begin{figure}
\centering
\begin{tikzpicture}[transform shape, scale=0.8, y=0.5cm]
	\node[dspnodeopen,dsp/label=left] (osc1) at (0,3) {$\cos\left(\omega_1 t + \theta_t^1\right)$};
	\node[coordinate] (cosc1) at (1,3) {};
	\node[coordinate] (cosc2) at (1,1) {};
	\node[dspnodeopen,dsp/label=left] (osc2) at (0,1) {$\cos\left(\omega_2 t + \theta_t^2\right)$};
	\node[dspmixer,dsp/label=left] (mul) at (1,2) {};
	\node[dspnodeopen,dsp/label=left] (osc3) at (0,-1) {$\cos\left(\omega_3 t + \theta_t^3\right)$};
	\node[coordinate] (cosc3) at (1,-1) {};
	\node[coordinate] (cosc4) at (1,-3) {};
	\node[dspnodeopen,dsp/label=left] (osc4) at (0,-3) {$\cos\left(\omega_4 t + \theta_t^4\right)$};
	\node[dspmixer,dsp/label=left] (mul2) at (1,-2) {};
	\node[coordinate] (cmul1) at (2,2) {};
	\node[coordinate] (cmul2) at (2,-2) {};
	\node[dspmixer,dsp/label=above] (mul3) at (2,0) {};
	\node[dspfilter] (hi_filt) at (4,0) {HPF};
	\node[dspnodeopen,dsp/label=right] (cout) at (6,0) {$\frac{1}{8}\cos\left( \omega t + \theta_t\right)$};
	\begin{scope}[start chain]
		\chainin (osc1);
		\chainin (cosc1) [join=by dspflow];
		\chainin (mul) [join=by dspline];
	\end{scope}
	\begin{scope}[start chain]
		\chainin (osc2);
		\chainin (cosc2) [join=by dspflow];
		\chainin (mul) [join=by dspline];
	\end{scope}
	\begin{scope}[start chain]
		\chainin (osc3);
		\chainin (cosc3) [join=by dspflow];
		\chainin (mul2) [join=by dspline];
	\end{scope}
	\begin{scope}[start chain]
		\chainin (osc4);
		\chainin (cosc4) [join=by dspflow];
		\chainin (mul2) [join=by dspline];
	\end{scope}
	\begin{scope}[start chain]
		\chainin (mul);
		\chainin (cmul1) [join=by dspflow];
		\chainin (mul3) [join=by dspline];
	\end{scope}
	\begin{scope}[start chain]
		\chainin (mul2);
		\chainin (cmul2) [join=by dspflow];
		\chainin (mul3) [join=by dspline];
		\chainin (hi_filt) [join=by dspflow];
		\chainin (cout) [join=by dspflow];
	\end{scope}
\end{tikzpicture}
\caption{Mixing stage designed to average 4 oscillators. At the output, $\omega = \sum_{i=1}^4 \omega_i$ and $\theta_t = \sum_{i=1}^4 \theta_t^i$.}
\label{fig:circuit4}
\end{figure}

With theoretical components, there is no limit to how many oscillators we may average in this fashion. As an example, Fig.\ref{fig:circuit4} shows the mixing stage with 4 oscillators.
Before the highpass filtering operation, and considering perfect mixers, the output would contain 3 terms whose frequency is close to DC, 4 terms close to $2f_c$ and one term---the term of interest---whose frequency is close to $4f_c$. The highpass filter in this case has to discriminate between $2f_c$ and $4f_c$.
\begin{figure}
\centering
\begin{tikzpicture}[transform shape, scale=0.8]
	\node[dspnodeopen,dsp/label=left] (osc1) at (0,0) {$\cos\left(\omega t + \theta_t\right)$};
	\node[dspmixer] (mul2) at (1,0) {};
	\node[dspfilter] (low_filt) at (3,0) {LPF};
	\node[dspfilter] (mul) at (2.25,-1) {Mul.};
	\node[dspfilter] (amp) at (3.75,-1) {Amp.};
	\node[dspnodefull] (splitter) at (5,0) {};
	\node[coordinate] (c1) at (5,-1) {};
	\node[coordinate] (c2) at (1,-1) {};
	\node[dspnodeopen,dsp/label=right] (cout) at (6,0) {$\frac{1}{2}\cos\left( \omega' t + \theta_t'\right)$};
	\begin{scope}[start chain]
		\chainin (osc1);
		\chainin (mul2) [join=by dspflow];
		\chainin (low_filt) [join=by dspline];
		\chainin (splitter) [join=by dspflow];
		\chainin (cout) [join=by dspflow];
	\end{scope}
	\begin{scope}[start chain]
		\chainin (splitter);
		\chainin (c1) [join=by dspline];
		\chainin (amp) [join=by dspflow];
		\chainin (mul) [join=by dspline];
		\chainin (c2) [join=by dspflow];
		\chainin (mul2) [join=by dspline];
	\end{scope}
\end{tikzpicture}
\caption{Frequency dividing stage. The Mul. block multiplies the frequency by $n-1$ \cite{Rubiola1992}.}
\label{fig:ndivider}
\end{figure}
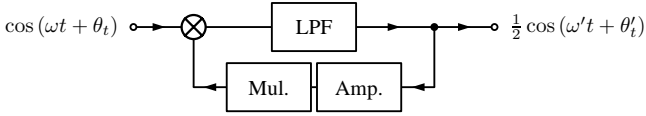

In the dividing stage, the $n$-frequency divider shown on Fig.\ref{fig:ndivider} is similar to the circuit presented on Fig.\ref{fig:circuit}---with an additional frequency multiplier in the feedback loop. Frequency multipliers and their possible drawbacks are described in detail in \cite{Rubiola1992}. The goal of the frequency multiplier is to take the system to the following steady-state equation:
\begin{equation}
	w' = w - (n-1)w' \qquad \theta_t' = \theta_t - (n-1)\theta_t'
\end{equation}
When the number of oscillators is a power of 2, stacking standard $2$-dividers is also an option; it can lead to circuits with lower phase noise overall \cite{jrilee2003}.

\subsection{Averaging the oscillator with itself}
\label{sec:delay_osc}
Improvements in phase noise performance may be extracted even when the original phase noise processes are not independent.
A typical case consists in averaging an oscillator with its own output delayed by some $\delta > 0$.
Assuming the original phase noise process decorrelates with time, the performance of the averaged oscillator should improve as $\delta$ increases.
Let the initial phase noise process $\theta_t$ follow the Wiener model of Sec.\ref{sec:model}. The averaged process is no longer a Wiener process; we can characterize the process and its power spectrum as follows.
\begin{proposition}
Let $\phi_t = (\theta_t + \theta_{t-\delta})/2$ be defined as the average of two delayed Wiener processes, and let $v_t = e^{\jmath \phi_t}$. The auto-correlation of $v_t$ is
\begin{equation}
	R_v(t, t+\tau) = \begin{cases}
		\exp \left(-\pi \beta \frac{|\tau|}{2}\right) \quad &|\tau| < \delta \\
		\exp \left(-\pi \beta \left(|\tau|-\frac{\delta}{2}\right)\right) \quad &|\tau| \geq \delta
	\end{cases}
	\label{eq:delayed_xcorr}
\end{equation}
and its PSD function is
\begin{equation}
	\begin{split}
		S_v(\omega) = &\frac{\exp \left(-\pi \beta \delta/2\right)}{(\pi \beta)^2 + \omega^2}\big(2 \pi \beta \cos(\omega\delta) - 2\omega \sin(\omega\delta)\big) \\
		-& \frac{\exp \left(-\pi \beta \delta/2\right)}{(\pi \beta/2)^2 + \omega^2}\big(\pi \beta \cos(\omega\delta) - 2\omega \sin(\omega\delta)\big) \\
		+& \frac{\pi \beta}{(\pi \beta/2)^2 + \omega^2}.
	\end{split}
\end{equation}
\end{proposition}
\begin{remark}
	Since $R_v(t, t+\tau)$ does not depend on $t$, $v_t$ is a stationary process. When $\delta$ is small, we recover the spectrum of the original process in \eqref{eq:psd}. When $\delta$ is large, the exponential terms vanish and the resulting spectrum corresponds to a Wiener process with half the variance of the original process.
\end{remark}
\begin{proof}
	We detail the proof in the appendix.
\end{proof} 

\section{Performance analysis}

\begin{figure}[t!]
		\scriptsize
		\begin{tikzpicture}
		\begin{axis}[
			width=\figurewidth,
			height=0.8\figurewidth,
			xmajorgrids,
			ymajorgrids,
			xlabel=Relative frequency from carrier (Hz),
			ylabel=Phase noise PSD (dBc/Hz),
			xmode=log,
			legend entries={
				Base oscillator,
				{Averaged, independent},
				{Averaged, $\delta=$1 $\mu$s},
				{Averaged, $\delta=$0.1 $\mu$s},
			},
			xmin=1e3,
			xmax=1e7,
			legend style={at={(0.01, 0.01)}, anchor=south west},
			legend cell align=left,
		]
			\addplot+ [color=col1, solid, mark=none] table {psd_log_base.data};
			\addplot+ [color=col2, solid, mark=o, mark repeat=25] table {psd_log_ind.data};
			\addplot+ [color=col6, solid, mark=+, mark repeat=25] table {psd_log_delta_1em6.data};
			\addplot+ [color=col4, solid, mark=x, mark repeat=25] table {psd_log_delta_1em7.data};
		\end{axis}
	\end{tikzpicture}
	\caption{PSD for an oscillator following the Wiener model, the oscillator averaged with itself as described in Sec.~\ref{sec:delay_osc} for different delay values. An average of two such independent oscillators using the circuit described in Sec.\ref{sec:base_circuit}.}
	\label{fig:psd}
\end{figure}
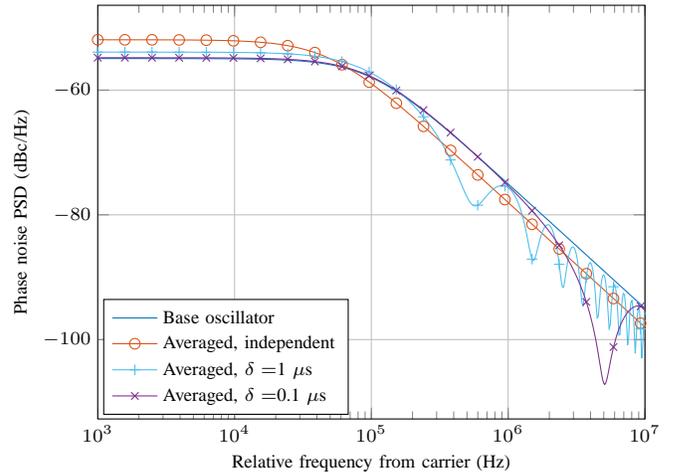

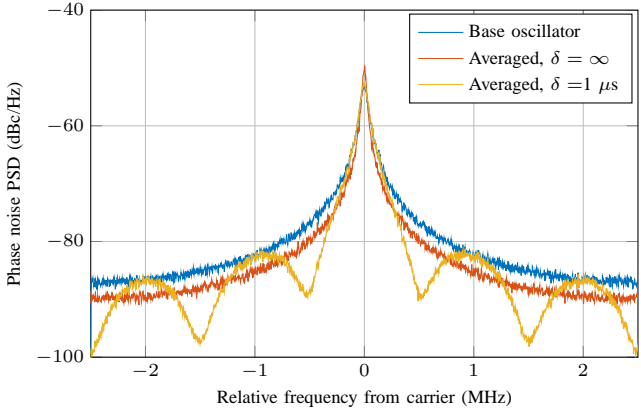
\begin{figure}[t!]
		\scriptsize
		\begin{tikzpicture}
		\begin{axis}[
			width=\figurewidth,
			height=0.7\figurewidth,
			xmajorgrids,
			ymajorgrids,
			xlabel=Relative frequency from carrier (MHz),
			ylabel=Phase noise PSD (dBc/Hz),
			legend entries={
				Base oscillator,
				{Averaged, $\delta=\infty$},
				{Averaged, $\delta=$1 $\mu$s},
			},
			xmin=-2.5,
			xmax=2.5,
			ymin=-100,
			ymax=-40,
			legend style={at={(0.99, 0.99)}, anchor=north east},
			legend cell align=left,
		]
			\addplot+ [color=col1, solid, mark=none] table {psd_lin_base.data};
			\addplot+ [color=col2, solid, mark=none, mark repeat=25] table {psd_lin_ind.data};
			\addplot+ [color=col3, solid, mark=none, mark repeat=25] table {psd_lin_delta_1em6.data};
		\end{axis}
	\end{tikzpicture}
	\caption{PSD for an oscillator following the Wiener model over a 5 MHz bandwith, an average of two such independent oscillators using the circuit described in Sec.\ref{sec:base_circuit}, and an oscillator averaged with itself with a delay of 1~$\mu$s.}
	\label{fig:lin_psd}
\end{figure}

In this section, we apply the results derived in this letter using practical values for the oscillator phase noise performances, e.g. from \cite{voicu2013} and \cite{cao2006}.
We plot first the PSD for a phase noise process following the Wiener model in Fig.\ref{fig:psd}; the basic oscillator used in this example is based on a voltage controlled oscillators with a wide tuning range---e.g. 0.1--65.8 GHz in \cite{Chen2009}.
We show the improvements obtained by averaging two independent oscillators using our basic circuit in Sec.\ref{sec:base_circuit}, and an oscillator averaged with delayed version of itself using the circuit in Sec.\ref{sec:delay_osc}.
Since larger delays decrease the correlation between the phase noise processes at the output of an oscillator, the performance of the independent oscillators may be obtained through the delay circuit by using a very large delay.
In both cases and under the Wiener model, the averaging circuit concentrates the phase noise oscillations closer to the carrier.
The phase noise is thus overall of lower frequency and thereby easier to track over time in most communication applications.
We see that the delay circuit distorts the PSD function at higher frequencies, and can concentrate phase noise effects closer to the carrier depending on the variance of the original oscillator and the length of the delay line.
Such a behavior is also reproduced on Fig.\ref{fig:lin_psd}, which shows the PSD over a 5 MHz bandwith.
The circuit greatly reduces the phase noise variance at periodic offsets from the carrier.
This may in turn improve the performance of multi-carrier systems operating with such oscillators.
Increasing the delay further reduces the period between the drops in Fig.\ref{fig:lin_psd}; the PSD function slowly tends to the independent case for larger delays.
\section{Conclusion}

In this letter, we described a circuit which averages out phase noise processes in local oscillators.
We showed that this circuit can be adapted to average multiple oscillators, as well as delayed versions of the same oscillator.
We provided an analysis of the phase noise process at the output, assuming that the input oscillators followed the Wiener model described in Sec.\ref{sec:model}.
In follow-up works, we plan to analyze the practical performance of such circuits.

In particular, we did not model the impact of non-idealities in the components on the resulting phase noise process.
Amplifiers are a major source of both amplitude noise and phase noise in oscillator designs \cite{Rubiola2008}.
Our additional amplifying step is a frequency divider, and analog frequency dividers lead to very low additional phase noise \cite{Ferre-Pikal1999, jrilee2003}
On the other hand, nonlinear mixers can decrease the performance of the averaging circuit and will leak harmonics of their strongest mixing modes in practical implementations.
This translate into frequency images around $2f_c$ in our basic circuit.
This effect has been demonstrated in part in \cite{llopis1993}.
These frequency images may be strongly reduced by additional filtering steps between the mixing operations, at the expense of increased amplitude noise.
A comprehensive analysis of the phase noise induced by mixers and overall $N$-frequency dividers can be found in \cite{Rubiola1992} and may provide guidelines to evaluate this trade-off.
The authors of \cite{Rubiola1992} also show that using the frequency dividing circuit at less-than-maximal amplitude provides additional gain in phase-noise reduction.

Finally, our analysis of the resulting phase noise process was also based on the steady-state solution of the system.
In practical systems, the output oscillation will go through a transient phase before reaching the steady state.
The duration of this transient phase in nonetheless low with state-of-the-art solutions---the settling time has been measured at less than 10 ns for a similar circuit \cite{Lee2006}.
Practical implementations are now necessary to assess whether this hold for the circuit presented in this work.
\appendix

\section{Derivation of the PSD for the oscillator averaged with itself}
\label{app:psd}
The auto-correlation of $v_t$ is defined as
\begin{equation}
	R_v(t, t + \tau) = \mathbb E\left[e^{\jmath\frac 12\left(\theta_t - \theta_{t+\tau} + \theta_{t-\delta} - \theta_{t + \tau - \delta}\right)}\right].
\end{equation}
For Wiener processes, $\theta_t - \theta_{t+\tau}$ follows a Gaussian distribution~\cite{lee2000b}. Here w compare two instances of the process on different time spans: $[t, t+\tau]$ and $[t-\delta, t - \delta + \tau]$. Assume first that $\tau > 0$.
\paragraph{If $\tau < \delta$} the process instances in the time spans $[t, t+\tau]$ and $[t-\delta, t - \delta + \tau]$ are independent. In this case, we have $R_v(t, t+\tau) = \mathbb E\left[e^{\jmath (T_1 + T_2)}\right]$, where $T_1$ and $T_2$ are 2 independent zero-mean Gaussian random variables both with variances $\pi \beta \tau/2$.
Identifying $R_v(t, t+\tau)$ as the characteristic function of a Gaussian random variable, we can write that $R_v(t, t+\tau) = \exp(-\frac 12\pi \beta \tau)$ when $\tau < \delta$.
\paragraph{If $\tau \geq \delta$} the process instances are not independent and they partly overlap. The process instances in the time spans $[t - \delta, t]$ and $[t + \tau - \delta, t + \tau]$ are independent; the process instances in the time span $[t, t+\tau-\delta]$ on the other hand are identical. In this case, $R_v(t, t+\tau) = \mathbb E\left[e^{\jmath (T_1 + T_2 + 2T_3)}\right]$ where $T_1$, $T_2$ and $T_3$ are zero-mean, independent Gaussian random variables representing the aforementioned process instances. Both $T_1$ and $T_2$ have variance $\pi \beta \delta/2$; $T_3$ has variance $\pi \beta (\tau - \delta)/2$. As before, we can conclude that $R_v(t, t+\tau) = \exp\left(- \pi \beta \left(\tau-\frac\delta 2\right) \right)$ when $\tau \geq \delta$.
The proof follows along the same lines when $\tau < 0$.
The auto-correlation is thus written as in \eqref{eq:delayed_xcorr}.

We now derive the PSD. Since the process is stationary, we can drop the time variable and write $R_v(t, t+\tau)$ as $R_v(\tau)$. The PSD is then
\begin{align}
	S_v(\omega) = \int_{-\infty}^\infty R_v(\tau) e^{\jmath \omega \tau} d\tau
\end{align}
The integral can be split in 4 parts, as in
\begin{align*}
    S_v(\omega) = &\int_{-\infty}^{-\delta} e^{\pi \beta (\tau+\delta/2)}e^{-\jmath \omega \tau} d\tau +\int_{-\delta}^0 e^{\pi \beta \tau/2}e^{-\jmath \omega \tau} d\tau \\
    + &\int_{0}^\delta e^{-\pi \beta \tau/2}e^{-\jmath \omega \tau} d\tau + \int_{\delta}^{\infty} e^{-\pi \beta (\tau-\delta/2)}e^{-\jmath \omega \tau} d\tau.
\end{align*}
Each integral can be solved analytically, leading to
\begin{align*}
	S_v(\omega) = &\frac{e^{\delta \pi\beta/2}}{\pi \beta - \jmath \omega}e^{-\delta(\pi \beta - \jmath \omega)} +\frac{1-e^{-\delta(\pi\beta/2 - \jmath \omega)}}{\pi\beta/2-\jmath \omega} \\
	+ &\frac{1-e^{-\delta(\pi\beta/2 + \jmath \omega)}}{\pi\beta/2+\jmath \omega}+\frac{e^{\delta \pi\beta/2}}{\pi \beta + \jmath \omega}e^{-\delta(\pi \beta + \jmath \omega)}
\end{align*}
Now, through basic Euler angle relations, we can group the relevant terms two-by-two and simplify them as
\begin{align*}
     &\frac{e^{\delta \pi\beta/2}}{\pi \beta - \jmath \omega}e^{-\delta(\pi \beta - \jmath \omega)}+\frac{e^{\delta \pi\beta/2}}{\pi \beta + \jmath \omega}e^{-\delta(\pi \beta + \jmath \omega)} \\
     =&\frac{e^{-\delta \pi\beta/2}}{(\pi \beta)^2 + \omega^2}\bigg(\pi\beta \left(e^{\jmath \omega\delta} + e^{-\jmath \omega\delta}\right) + \jmath\omega\left(e^{\jmath \omega\delta} - e^{-\jmath \omega\delta}\right)\bigg) \\
     =&\frac{e^{-\delta \pi\beta/2}}{(\pi \beta)^2 + \omega^2}\bigg(2 \pi \beta \cos(\omega\delta) - 2\omega \sin(\omega\delta)\bigg)
\end{align*}
Proceeding similarly for the remaining terms and summing them results in \eqref{eq:delayed_xcorr}.


\end{document}